\documentclass[letters,a4paper,fleqn,usenatbib]{mnras}
\usepackage[T1]{fontenc}
\usepackage{ae,aecompl}
\usepackage{graphicx}	
\usepackage{amsmath}	
\usepackage{amssymb}	
\renewcommand{\vec}[1]{\mbox{\boldmath $#1$}}
\title[Modelling differential rotation of red giants] 
{Modelling differential rotation of red giants: the case of the evolved sun}
\author[L.~Kitchatinov \& A.~Nepomnyashchikh]
{Leonid~Kitchatinov\thanks{E-mail: kit@iszf.irk.ru}
and Alexander~Nepomnyashchikh
\\
Institute of Solar-Terrestrial Physics, Lermontov Str. 126A, 664033, Irkutsk, Russia
}

\date{Accepted XXX. Received YYY; in original form ZZZ}

\pubyear{2019}
\begin{document}
\label{firstpage}
\pagerange{\pageref{firstpage}--\pageref{lastpage}}
\maketitle

\begin{abstract}
Asteroseismology has revealed that cores of red giants rotate about one order of
magnitude faster than their convective envelopes. This paper attempts an explanation for this rotational state in terms of the theory of angular momentum transport in stellar convection zones. A differential rotation model based on the theory is applied to a sequence of evolutionary states of a red giant of one solar mass. The model computations show a rotation of about ten times faster in the cores compared to the stellar surface. This rotational state is caused by the non-diffusive downward convective transport of angular momentum. The contrast in rotational rates between core and envelope increases with the radius (age) of the star. Seismologically detected scaling for the spindown of the giants' cores is also reproduced.
\end{abstract}

\begin{keywords}
hydrodynamics -- stars: rotation -- stars: interiors -- Sun: evolution
\end{keywords}

\section{Introduction}
Asteroseismology of red giants has revealed a strong inhomogeneity in the rotation of their interiors. \citet{Bea12} analysed the frequency splitting of global oscillations of three red giants 20 to 50 per cent more massive than the Sun and concluded that the cores of the stars rotate about ten times faster than their extended convective envelopes. \citet{Dea12} detected a similar rotational state in a giant of subsolar mass. The extensive  statistics of seismological detections by \citet{Mea12} confirmed the relatively rapid rotation of giants' core and established scaling for the core spindown during the red giant phase.

This paper suggests an explanation for the seismologically detected rotation of red giants in the framework of the differential rotation theory. The explanation may seem to be obvious in view of the core contraction, envelope expansion and angular momentum conservation in the course of a star ascendance on the red giant branch (RGB). Quantitative realisation of this idea met problems however.

\citet{Cea13} applied the stellar evolution model by \citet{Eea10} to the {\it Kepler} target KIC~7341231 whose rotational state was detected by \citet{Dea12}. The resulting rotation profile was too steep compared to its seismological detection. \citet{Cea13} concluded that angular momentum transport by the meridional flow and the rotational shear mixing of the standard shellular rotation models do not suffice to explain the differential rotation of the red giant and some additional mechanism for the transport of angular momentum is needed \citep[cf., however,][]{FPJ19}. \cite{Eea19} confirmed the necessity for an efficient mechanism complementary to eddy viscosity. More specifically, angular momentum transport proportional to the rotational shear (viscous transport) cannot reproduce seismological data on the rotation of subgiants and red giants simultaneously.

In this paper, we argue that allowance for angular momentum transport by convective turbulence can help in resolving the problem. The ability of rotating turbulence to transport angular momentum even in the case of uniform background rotation has long been recognized \citep{L41}. Non-diffusive transport is conventionally named the \lq$\Lambda$-effect' \citep{R89}. The effect results from an anisotropy of turbulent mixing. Predominantly radial convective mixing produces a downward increase in angular velocity due to the tendency towards angular momentum conservation by mixed fluid parcels. A hypothetical turbulence in stellar radiation zones can produce the $\Lambda$-effect as well, but predominantly horizontal mixing in stably stratified fluids transports angular momentum upward in radius \citep{KB12}. Differential rotation in stellar convection zones can be to some extent understood as the balance between the $\Lambda$-effect and eddy viscosity. The direct numerical simulations of \citet{BP09} have shown a strong downward increase in rotation rate
in the lower part of the convection zones of red giants.

This paper concerns the evolution of the structure and differential rotation for the red giant stage of a star of one solar mass. The choice of 1M$_{\sun}$ star is motivated by valuable guidance provided by helioseismology and by recent data on the rotation of solar-type stars \citep{Mea16,vSea19} for this case. Helioseismology revealed in particular the surface shear layer in the upper part of the convective envelope where the rotation rate increases sharply with depth \citep{Tea96,BSG14}. The surface shear layer in the Sun and large rotational shear in extended convection zones of red giants can be of the same origin \citep{K16}.
\section{Method and model}
Modelling differential rotation requires input information on the stellar structure.
The structure evolution for a star of 1M$_{\sun}$ and metallicity $Z=0.02$ was computed with the code {\scriptsize MESA} by \citet{Pea11}, version {\scriptsize 11532}\footnote{\url{http://mesa.sourceforge.net}.}. The results to follow refer to the stages of the star ascendance and evolution on the RGB when the star's radius increases from 1.5 to 15 $R_{\sun}$. This range of radius corresponds to ages between 10.8 and 12.3 Gyr, ending shortly before the onset of Helium burning.

\begin{figure}
	\includegraphics[width=\columnwidth]{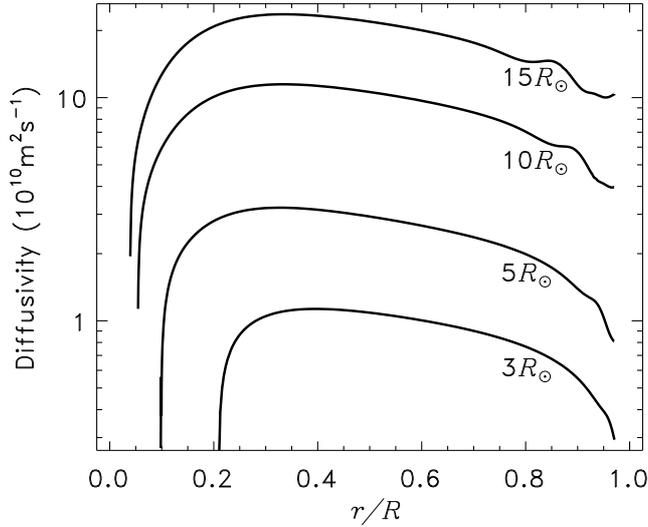}
    \caption{Profiles of the eddy diffusivity of Eq.\,(\ref{1}) for several
            evolutionary states of the 1M$_{\sun}$ red giant. The curves are marked by the corresponding stellar radius.
        }
    \label{f1}
\end{figure}

Evolved stars have extended convection zones. The intensity and characteristic time of convective mixing can be estimated in terms of the eddy diffusivity
\begin{equation}
    \nu_{_\mathrm{T}} = \ell v_\mathrm{c}/3,
    \label{1}
\end{equation}
where $\ell$ is the mixing-length and $v_\mathrm{c}$ is the convective velocity, both provided by the {\scriptsize MESA} code ($\alpha_{_\mathrm{MLT}} =\ell/H_\mathrm{p} = 2$ in the computations of this paper). Figure\,\ref{f1} shows the eddy diffusivity profiles for several values of the evolving star radius. The radius $R$ increases with age.

\begin{figure}
	\includegraphics[width=\columnwidth]{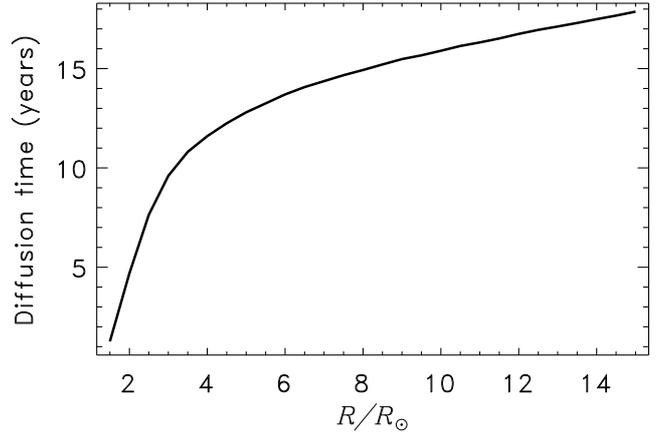}
    \caption{Diffusion time of Eq.\,(\ref{2}) as a function of the evolving star radius.
        }
    \label{f2}
\end{figure}

The diffusivity of Fig.\,\ref{f1} increases with $R$, so that the diffusion time
\begin{equation}
    T_\mathrm{d} = \left( \int\limits_{r_\mathrm{i}}^{r_\mathrm{e}} \frac{\mathrm{d}r}{\sqrt{\nu_{_\mathrm{T}}}} \right)^2 ,
    \label{2}
\end{equation}
varies moderately and remains in the order of one decade. In this equation, $r_\mathrm{i}$ is the inner radius of the convection zone and $r_\mathrm{e} = 0.97 R$ is the external radius of the computation domain for differential rotation modelling. The $\Lambda$-effect scales with the eddy viscosity of Eq.\,(\ref{1}) \citep{R89}. Therefore, $T_\mathrm{d}$ in Eq.\,(\ref{2}) is also the characteristic time for establishing  equilibrium between the $\Lambda$-effect and eddy viscosity. The diffusion time of Fig.\,\ref{f2} is short compared to evolutionary time-scales. The differential rotation can therefore be computed with a steady model.

The differential rotation was computed with the mean-field model of \citet{KO11,KO12}. The model closely reproduces seismological data on the internal rotation of the Sun and the increasing trend of surface differential rotation with the stellar temperature observed by \citet{Bea05} and \citet{BA16}. The differential rotation model computes the angular velocity distribution inside the convection zone only. It does not apply to the radiation zone. It is known from helioseismology, however, that the bulk of the radiation zone rotates almost rigidly with the rate close to the (latitude-averaged) rotation rate of the base of the convection zone \citep{Sea98}. The radiation core and convection zone are therefore linked in a spinning-down star. The origin of the link is currently uncertain. Coupling by Maxwell stress of an internal magnetic field is a likely possibility that can simultaneously explain rigid rotation in the bulk of the radiation zone and a slender tachocline on its top \citep{RK96,MC99}.

Whatever the origin of the link is, we assume that it is active during the late evolution of solar-type stars so that the region beneath the convection zone
still rotates rigidly with the latitude-averaged angular velocity $\overline\Omega (r_\mathrm{i})$ of the inner boundary of the convection zone. The latitude-averaging,
\begin{equation}
    \overline{\Omega}(r) = \frac{3}{4}\int\limits_{-\pi/2}^{\pi/2}
    \Omega(r,\lambda)\cos^3\lambda\,\mathrm{d}\lambda ,
    \label{3}
\end{equation}
is the \lq angular momentum correspondent': shellular rotation with angular velocity $\overline{\Omega}(r)$ has the same angular momentum as the differential rotation  $\Omega (r,\lambda)$ dependent on the latitude $\lambda$. Asteroseismology is currently not certain about the rotation profile in radiative interiors of red giants. \citet{Dea16} detected rigid rotation in the helium core of a giant of slightly super-solar mass but found indications of an outward decrease in rotation rate in the hydrogen burning shell just beneath the convection zone \citep[cf. also][]{Dea14}. We return to discussing the consequences of the assumed rigid rotation of the radiation zone later.

The angular momentum of a star has to be specified in order to model the differential rotation. The rotation rate of solar-type stars is known to decrease with age. \citet{Mea16} and \citet{vSea19} have shown that the spindown essentially stops when the Rossby number of a star increases to the value of about 2 \citep[see also][]{R84}. The termination of the spindown can be explained by the switch-off of the global dynamo at sufficiently slow rotation \citep{KN17}. The estimations of that paper suggest that the star of 1M$_{\sun}$ reaches the state of marginal dynamo at the age of about 5\,Gyr and its total angular momentum
\begin{equation}
    M = \frac{8\pi}{3}\int\limits_{0}^{R} \rho(r)\overline{\Omega}(r) r^4\,\mathrm{d}r
    \label{4}
\end{equation}
equals $M_\mathrm{thr} = 1.88\times10^{41}$~kg\,m$^2$\,s$^{-1}$ at this age. We neglect the angular momentum loss in the further evolution of the star and
adjust the mean rotation rate $\overline{\Omega}(R)$ on the top boundary, which is an input parameter of our model, with the condition for the total angular momentum to equal the above constant value of $M_\mathrm{thr}$.

\begin{figure}
	\includegraphics[width=\columnwidth]{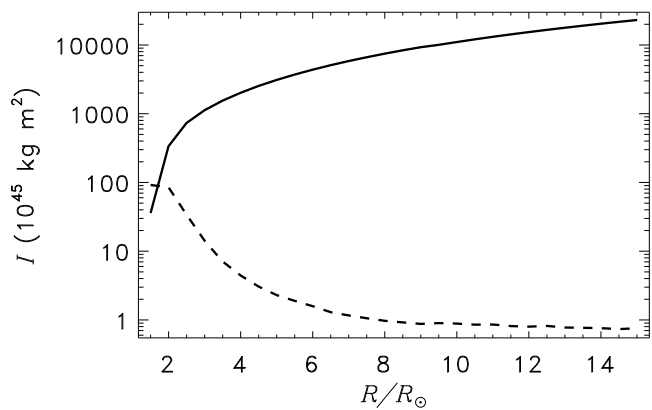}
    \caption{Momenta of inertia of the convection zone (full line) and the
            radiative interior (dashed) as function of the radius $R$ of the evolving star.
        }
    \label{f3}
\end{figure}

Figure~\ref{f3} shows the rotational inertia
\begin{equation}
    I = \frac{8\pi}{3}\int\limits_{r_1}^{r_2} \rho(r) r^4\,\mathrm{d}r
    \label{5}
\end{equation}
of the core and convective envelope as a function of the radius $R$ of the star ($r_1$ and $r_2$ equal $0$ and $r_\mathrm{i}$ for the core and $r_\mathrm{i}$ and $R$ for the envelope, respectively). The core can comprise a considerable part of the angular momentum at the beginning of the star ascendance on the RGB only. Later on, almost all angular momentum belongs to the convection zone. Therefore, the assumption of rigid rotation in the radiative interior affects the computed differential rotation of the convection zone only slightly. However, the assumption is consequential for the estimated contrast in rotation rate between the core and the surface. In this sense, our model can be understood as estimating only those parts of the differential rotation which is comprised by the convection zone alone.

The 2D differential rotation model solves jointly the equations for angular velocity, meridional flow and heat transport as functions of radius and latitude
in a stellar convection zone. Three modifications of the version of the model by \citet{KO11} were necessary to adjust it for application to the red giants:
\begin{enumerate}
\item The most consequential modification is the neglect of the advection term $\rho T \vec{V}\cdot\vec{\nabla}S$ in the heat transport equation ($\vec{V}$ is the meridional flow velocity and $S$ is the specific entropy). The neglect of the entropy advection is dictated by an internal contradiction in mixing-length formalism. The formalism underestimates eddy transport coefficients in low-density regions of convection zones. The underestimation can result in an instability to large-scale modes of thermal convection \citep{Tea94}. Our model is unstable when applied to red giants. The instability is switched off by neglect of the entropy advection.
\item The anisotropy parameter $a$ \citep[cf. eq. (A1) in][]{KO11} was increased from 2 to 3. This parameter is important for differential rotation in those regions of a star where the Coriolis number
    \begin{equation}
        \Omega^* = 2\tau\Omega
        \label{7}
    \end{equation}
    is small ($\tau = \ell/v_\mathrm{c}$ is the convective turnover time). This is the case with the near-surface region of the Sun in particular. \citet{BSG14} found that the normalized rotational shear $\frac{r}{\Omega}\frac{\partial\Omega}{\partial r}$ near the solar surface is constant with latitude. The constancy can be explained in terms of the $\Lambda$-effect for the case of a small Coriolis number \citep{K16}. The value of -1 by \citet{BSG14} for the normalized shear is reproduced with the anisotropy parameter $a = 3$. This is why we use this value for giants where the Coriolis number can be small as well.
\item Background stratification in convection zones of main-sequence stars was formerly approximated by integrating the corresponding equations downward from the top boundary. This procedure can result in error accumulation when applied to extended convection zones of red giants. The background stratification of density, temperature and other input parameters of the differential rotation model is now taken directly from the {\scriptsize MESA} model for the stellar structure.
\end{enumerate}

\begin{figure}
	\includegraphics[width=\columnwidth]{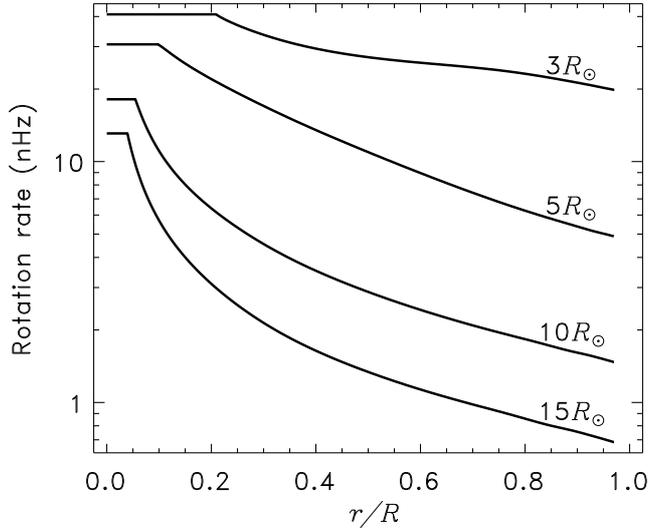}
    \caption{Profiles of the internal rotation rate of 1M$_{\sun}$
            star for several instants of its evolution on the RGB.
            The curves are marked by the corresponding stellar radius.
        }
    \label{f4}
\end{figure}

\section{Results and discussion}
Figure\,\ref{f4} shows the radial profiles of the rotation rate $\nu(r) = \overline{\Omega}(r)/(2\pi)$ computed with our model ($\overline{\Omega}(r)$ is the latitude-averaged angular velocity of Eq.\,(\ref{3})). In all cases shown, the core rotates faster than the convective envelope. The rotation rate varies smoothly with position so that the rotation rate of the envelope cannot be uniquely defined. We use the surface rotation rate $\overline{\Omega}(R)$ as a potentially observable parameter and the mean angular velocity
\begin{equation}
    \Omega_\mathrm{env} = M_\mathrm{env}/I_\mathrm{env}
    \label{8}
\end{equation}
to characterise the envelope's rate of rotation. In this equation, $M_\mathrm{env}$ is the envelope's angular momentum of Eq.\,(\ref{4}), where the low limit of the integration is changed to $r_\mathrm{i}$, and $I_\mathrm{env}$ is the envelope's moment of inertia of Eq.\,(\ref{5}).

\begin{figure}
	\includegraphics[width=\columnwidth]{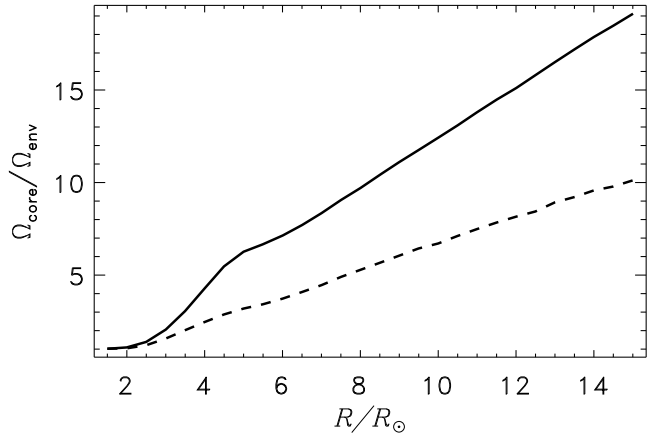}
    \caption{Ratio of the core to envelope rotation rates. The full line
            shows the ratio for the surface value of $\overline{\Omega}(R)$
            for the envelope rotation and the dashed line - for the mean angular velocity of Eq.\,(\ref{8}).
        }
    \label{f5}
\end{figure}

Figure\,\ref{f5} shows the core-to-envelope ratio of rotation rates as the function of the stellar radius. The ratio is smaller for the rate of Eq.\,(\ref{8}) but the characteristic value of the ratio of order 10 agrees with asteroseismological detections \citep{Bea12,Dea12,Mea12} for either definition of the envelope's rate.
Figure\,\ref{f5} however underestimates by a factor of 0.4 the rotational contrast of about 11 detected by \citet{Dea16} for the near-solar mass giant KIC\,4448777. If confirmed, the underestimation indicates a substantial contribution to the rotational contrast by the radiative interior missed in our model.

\begin{figure}
	\includegraphics[width=\columnwidth]{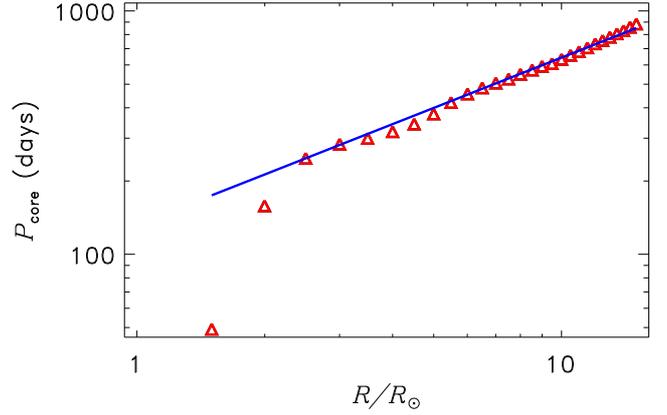}
    \caption{Triangles show the rotation periods of the core
            computed for an evolutionary series of stellar structure of different radii. The full line is the power law approximation of Eq.\,(\ref{9}), $p = 0.687$.
        }
    \label{f6}
\end{figure}

The core rate of Fig.\,\ref{f4} is smaller for stars of greater age. \citet{Mea12} approximated the core spindown by the power law
\begin{equation}
    P_\mathrm{rot} \propto R^p ,
    \label{9}
\end{equation}
where $P_\mathrm{rot}$ is the rotation period. They found the power index $p = 0.7 \pm 0.3$ for the RGB phase. The core rates of our computations are close to the power law of Eq.\,(\ref{9}) for $R > 2.5R_{\sun}$ only (Fig.\,\ref{f6}). The power index $p = 0.69$ of the least-square fit for this range agrees with the seismological results \citep[cf. eq.\,(26) of][]{Mea12}. The decrease in the core rotation rate in our model is a consequence of the increasing rotational inertia of the expanding star (Fig.\,\ref{f3}).

\begin{figure}
	\includegraphics[width=\columnwidth]{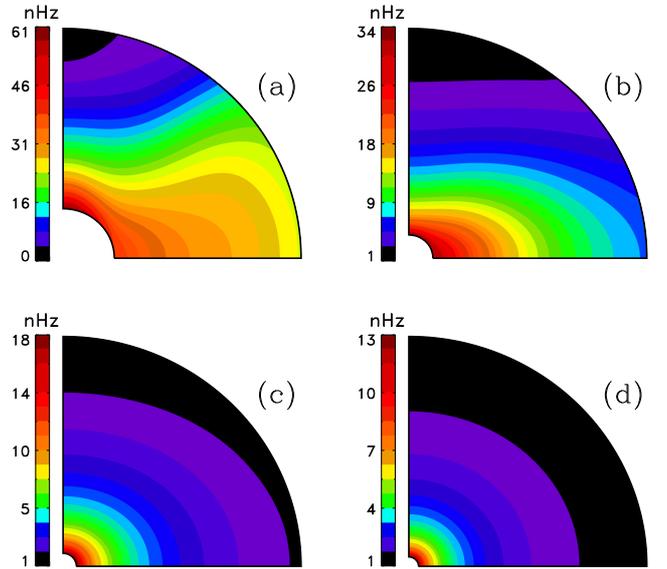}
    \caption{Rotation rate isolines in the north-west quadrants of meridional
            cross-sections of convection zones. Parts (a), (b), (c) and (d) correspond to stellar radii of 3, 5, 10 and 15 $R_{\sun}$ respectively.
        }
    \label{f7}
\end{figure}

\begin{figure}
	\includegraphics[width=\columnwidth]{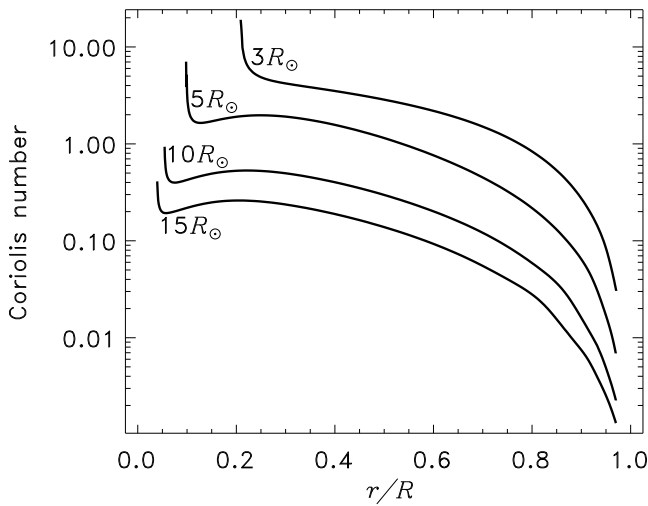}
    \caption{Profiles of the Coriolis number of Eq.\,(\ref{7}) estimated with the
            latitude-averaged angular velocity of Eq.\,(\ref{3}) for the same models as in Fig.\,\ref{f7}. The curves are marked by the corresponding stellar radius.
        }
    \label{f8}
\end{figure}

Figure\,\ref{f7} shows that the red giants can possess latitudinal differential rotation in line with the strong radial shear. The decrease in angular velocity with latitude is quite pronounced in computations for the early red giant phase. The computed differential rotation changes from latitude-dependent to shellular structure as the star ages. This change can be interpreted in terms of the $\Lambda$-effect  dependence on the Coriolis number. The non-diffusive flux of angular momentum changes from the radial inward direction to the equator-ward and parallel to the rotation axis as the Coriolis number increases \citep[cf. fig.\,3 in][]{K13,K19}. Fluxes of either direction result in an increase in the rotation rate with depth but the radial inward transport is more efficient in producing this rotation inhomogeneity. The axial transport produces the equatorial acceleration seen in parts (a) and (b) of Fig.\,\ref{f7}. The Coriolis number profiles in Fig.\,\ref{f8} support this interpretation. The profiles corresponding to the shellular rotation of parts (c) and (d) in Fig.\,\ref{f7} are smaller than one.

The power index $n$ in the best fit of convection zone rotation of Fig.\,\ref{f4} by the power law $\overline{\Omega}(r) \propto r^{-n}$ varies as $n = 0.47, 0.80, 0.87, 0.92$ for $R/R_{\sun} = 3, 5, 10, 15$, respectively. For the smallest Coriolis number of our computations, the power index thus approaches the value of $n = 1$ detected by \citet{BSG14} for the surface shear layer of the Sun.

Most red giants in seismological studies rotate faster than the star in our model. This is because one solar mass is the low bound of the mass range in the studies \citep{Mea12,Tea17,Eea19}. The rotation rate of main-sequence progenitors of the red giants increases with mass \citep[cf. fig.\,2 in][]{K67}. The angular momentum of the contemporary Sun does not allow a faster rotation at the red giant stage compared to our model.
\section{Conclusion}
The asteroseismologically detected faster rotation of the red giants' cores compared to their extended convective envelopes can be at least partly explained by the non-diffusive downward transport of angular momentum by turbulent convection known as the $\Lambda$-effect. The explanation is quantitatively confirmed with the model of stellar differential rotation applied to a red giant of one solar mass.
The explanation also suggests that the large radial shear in the rotation of the red giants has the same origin as the surface shear layer in the Sun.
\section*{Acknowledgements}
The authors are thankful to Dr Georges Meynet for pertinent and constructive comments.
This work was supported by the Russian Foundation for Basic Research (project 17-02-00016) and by budgetary funding of the Basic Research Program II.16.
\bibliographystyle{mnras}
\bibliography{paper}

\begin{thebibliography}{}
\makeatletter
\relax
\def\mn@urlcharsother{\let\do\@makeother \do\$\do\&\do\#\do\^\do\_\do\%\do\~}
\def\mn@doi{\begingroup\mn@urlcharsother \@ifnextchar [ {\mn@doi@}
  {\mn@doi@[]}}
\def\mn@doi@[#1]#2{\def\@tempa{#1}\ifx\@tempa\@empty \href
  {http://dx.doi.org/#2} {doi:#2}\else \href {http://dx.doi.org/#2} {#1}\fi
  \endgroup}
\def\mn@eprint#1#2{\mn@eprint@#1:#2::\@nil}
\def\mn@eprint@arXiv#1{\href {http://arxiv.org/abs/#1} {{\tt arXiv:#1}}}
\def\mn@eprint@dblp#1{\href {http://dblp.uni-trier.de/rec/bibtex/#1.xml}
  {dblp:#1}}
\def\mn@eprint@#1:#2:#3:#4\@nil{\def\@tempa {#1}\def\@tempb {#2}\def\@tempc
  {#3}\ifx \@tempc \@empty \let \@tempc \@tempb \let \@tempb \@tempa \fi \ifx
  \@tempb \@empty \def\@tempb {arXiv}\fi \@ifundefined
  {mn@eprint@\@tempb}{\@tempb:\@tempc}{\expandafter \expandafter \csname
  mn@eprint@\@tempb\endcsname \expandafter{\@tempc}}}

\bibitem[\protect\citeauthoryear{{Balona} \& {Abedigamba}}{{Balona} \&
  {Abedigamba}}{2016}]{BA16}
{Balona} L.~A.,  {Abedigamba} O.~P.,  2016, \mnras, 461, 497

\bibitem[\protect\citeauthoryear{{Barekat}, {Schou}  \& {Gizon}}{{Barekat}
  et~al.}{2014}]{BSG14}
{Barekat} A.,  {Schou} J.,   {Gizon} L.,  2014, \aap, 570, L12

\bibitem[\protect\citeauthoryear{{Barnes}, {Collier Cameron}, {Donati},
  {James}, {Marsden}  \& {Petit}}{{Barnes} et~al.}{2005}]{Bea05}
{Barnes} J.~R.,  {Collier Cameron} A.,  {Donati} J.-F.,  {James} D.~J.,
  {Marsden} S.~C.,   {Petit} P.,  2005, MNRAS, 357, L1

\bibitem[\protect\citeauthoryear{{Beck} et~al.,}{{Beck} et~al.}{2012}]{Bea12}
{Beck} P.~G.,  et~al., 2012, \nat, 481, 55

\bibitem[\protect\citeauthoryear{{Brun} \& {Palacios}}{{Brun} \&
  {Palacios}}{2009}]{BP09}
{Brun} A.~S.,  {Palacios} A.,  2009, \apj, 702, 1078

\bibitem[\protect\citeauthoryear{{Ceillier}, {Eggenberger}, {Garc{\'{\i}}a}  \&
  {Mathis}}{{Ceillier} et~al.}{2013}]{Cea13}
{Ceillier} T.,  {Eggenberger} P.,  {Garc{\'{\i}}a} R.~A.,   {Mathis} S.,  2013,
  \aap, 555, A54

\bibitem[\protect\citeauthoryear{{Deheuvels} et~al.,}{{Deheuvels}
  et~al.}{2012}]{Dea12}
{Deheuvels} S.,  et~al., 2012, \apj, 756, 19

\bibitem[\protect\citeauthoryear{{Deheuvels} et~al.,}{{Deheuvels}
  et~al.}{2014}]{Dea14}
{Deheuvels} S.,  et~al., 2014, \aap, 564, A27

\bibitem[\protect\citeauthoryear{{Di Mauro} et~al.,}{{Di Mauro}
  et~al.}{2016}]{Dea16}
{Di Mauro} M.~P.,  et~al., 2016, \apj, 817, 65

\bibitem[\protect\citeauthoryear{{Eggenberger}, {Miglio}, {Montalban},
  {Moreira}, {Noels}, {Meynet}  \& {Maeder}}{{Eggenberger}
  et~al.}{2010}]{Eea10}
{Eggenberger} P.,  {Miglio} A.,  {Montalban} J.,  {Moreira} O.,  {Noels} A.,
  {Meynet} G.,   {Maeder} A.,  2010, \aap, 509, A72

\bibitem[\protect\citeauthoryear{{Eggenberger} et~al.,}{{Eggenberger}
  et~al.}{2019}]{Eea19}
{Eggenberger} P.,  et~al., 2019, \aap, 621, A66

\bibitem[\protect\citeauthoryear{{Fuller}, {Piro}  \& {Jermyn}}{{Fuller}
  et~al.}{2019}]{FPJ19}
{Fuller} J.,  {Piro} A.~L.,   {Jermyn} A.~S.,  2019, \mnras, 485, 3661

\bibitem[\protect\citeauthoryear{{K{\"a}pyl{\"a}}}{{K{\"a}pyl{\"a}}}{2019}]{K19}
{K{\"a}pyl{\"a}} P.~J.,  2019, \aap, 622, A195

\bibitem[\protect\citeauthoryear{{Kitchatinov}}{{Kitchatinov}}{2013}]{K13}
{Kitchatinov} L.~L.,  2013, in {Kosovichev} A.~G.,  {de Gouveia Dal Pino} E.,
  {Yan} Y.,  eds,  IAU Symposium Vol. 294, Solar and Astrophysical Dynamos and
  Magnetic Activity. pp 399--410

\bibitem[\protect\citeauthoryear{{Kitchatinov}}{{Kitchatinov}}{2016}]{K16}
{Kitchatinov} L.~L.,  2016, Astronomy Letters, 42, 339

\bibitem[\protect\citeauthoryear{{Kitchatinov} \& {Brandenburg}}{{Kitchatinov}
  \& {Brandenburg}}{2012}]{KB12}
{Kitchatinov} L.~L.,  {Brandenburg} A.,  2012, Astronomische Nachrichten, 333,
  230

\bibitem[\protect\citeauthoryear{{Kitchatinov} \&
  {Nepomnyashchikh}}{{Kitchatinov} \& {Nepomnyashchikh}}{2017}]{KN17}
{Kitchatinov} L.,  {Nepomnyashchikh} A.,  2017, \mnras, 470, 3124

\bibitem[\protect\citeauthoryear{{Kitchatinov} \& {Olemskoy}}{{Kitchatinov} \&
  {Olemskoy}}{2011}]{KO11}
{Kitchatinov} L.~L.,  {Olemskoy} S.~V.,  2011, \mnras, 411, 1059

\bibitem[\protect\citeauthoryear{{Kitchatinov} \& {Olemskoy}}{{Kitchatinov} \&
  {Olemskoy}}{2012}]{KO12}
{Kitchatinov} L.~L.,  {Olemskoy} S.~V.,  2012, \mnras, 423, 3344

\bibitem[\protect\citeauthoryear{{Kraft}}{{Kraft}}{1967}]{K67}
{Kraft} R.~P.,  1967, \apj, 150, 551

\bibitem[\protect\citeauthoryear{{Lebedinskii}}{{Lebedinskii}}{1941}]{L41}
{Lebedinskii} A.~I.,  1941, \azh, 18, 10

\bibitem[\protect\citeauthoryear{{MacGregor} \& {Charbonneau}}{{MacGregor} \&
  {Charbonneau}}{1999}]{MC99}
{MacGregor} K.~B.,  {Charbonneau} P.,  1999, \apj, 519, 911

\bibitem[\protect\citeauthoryear{{Metcalfe}, {Egeland}  \& {van
  Saders}}{{Metcalfe} et~al.}{2016}]{Mea16}
{Metcalfe} T.~S.,  {Egeland} R.,   {van Saders} J.,  2016, \apjl, 826, L2

\bibitem[\protect\citeauthoryear{{Mosser} et~al.,}{{Mosser}
  et~al.}{2012}]{Mea12}
{Mosser} B.,  et~al., 2012, \aap, 548, A10

\bibitem[\protect\citeauthoryear{{Paxton}, {Bildsten}, {Dotter}, {Herwig},
  {Lesaffre}  \& {Timmes}}{{Paxton} et~al.}{2011}]{Pea11}
{Paxton} B.,  {Bildsten} L.,  {Dotter} A.,  {Herwig} F.,  {Lesaffre} P.,
  {Timmes} F.,  2011, \apjs, 192, 3

\bibitem[\protect\citeauthoryear{{Rengarajan}}{{Rengarajan}}{1984}]{R84}
{Rengarajan} T.~N.,  1984, ApJ, 283, L63

\bibitem[\protect\citeauthoryear{{R\"udiger}}{{R\"udiger}}{1989}]{R89}
{R\"udiger} G.,  1989, {Differential Rotation and Stellar Convection. Gordon \&
  Breach, New York}

\bibitem[\protect\citeauthoryear{{R\"udiger} \& {Kitchatinov}}{{R\"udiger} \&
  {Kitchatinov}}{1996}]{RK96}
{R\"udiger} G.,  {Kitchatinov} L.~L.,  1996, \apj, 466, 1078

\bibitem[\protect\citeauthoryear{{Schou} et~al.,}{{Schou} et~al.}{1998}]{Sea98}
{Schou} J.,  et~al., 1998, \apj, 505, 390

\bibitem[\protect\citeauthoryear{{Thompson} et~al.,}{{Thompson}
  et~al.}{1996}]{Tea96}
{Thompson} M.~J.,  et~al., 1996, Science, 272, 1300

\bibitem[\protect\citeauthoryear{{Triana}, {Corsaro}, {De Ridder}, {Bonanno},
  {P{\'e}rez Hern{\'a}ndez}  \& {Garc{\'{\i}}a}}{{Triana} et~al.}{2017}]{Tea17}
{Triana} S.~A.,  {Corsaro} E.,  {De Ridder} J.,  {Bonanno} A.,  {P{\'e}rez
  Hern{\'a}ndez} F.,   {Garc{\'{\i}}a} R.~A.,  2017, \aap, 602, A62

\bibitem[\protect\citeauthoryear{{Tuominen}, {Brandenburg}, {Moss}  \&
  {Rieutord}}{{Tuominen} et~al.}{1994}]{Tea94}
{Tuominen} I.,  {Brandenburg} A.,  {Moss} D.,   {Rieutord} M.,  1994, \aap,
  284, 259

\bibitem[\protect\citeauthoryear{{van Saders}, {Pinsonneault}  \&
  {Barbieri}}{{van Saders} et~al.}{2019}]{vSea19}
{van Saders} J.~L.,  {Pinsonneault} M.~H.,   {Barbieri} M.,  2019, \apj, 872,
  128

\makeatother
\end{thebibliography}
\bsp	
\label{lastpage}
\end{document}